\documentclass[twocolumn,twoside,slac_two]{revtex4}
\usepackage{graphicx}
\usepackage{fancyhdr}
\begin{document}

\title{Charmless hadronic $B$ and $B_s$ decays in perturbative QCD approach }


\author{Cai-Dian L\"u}

\affiliation{Institute of High Energy
Physics, CAS, P.O.Box 918(4), Beijing 100049, China and\\
Theoretical Physics Center for Science Facilities (TPCSF), CAS,
Beijing 100049, China }

\begin{abstract}
 We review the  perturbative QCD approach
study of hadronic B decays. Utilizing the constrained parameters in
these well measured decay channels, we study most of the possible
charmless $B_s \to PP$, $PV$ and $VV$ decay channels in the
perturbative QCD approach. In addition to the branching ratios and
CP asymmetries, we also give predictions to the polarization
fractions of the vector meson final states. The size of SU(3)
breaking effect is also discussed. These predictions can be tested
by the future   LHCb experiment.
%\keywords{$B_s$ decays; factorization; pQCD.}
\end{abstract}

\maketitle

\thispagestyle{fancy}


%\catchline{}{}{}{}{}
%
%%%%%%%%%%%%%%%%%%%%%%%%%%%%%%%%%%%%%%%%%%%%%%%%%%%%%%%%%%%%%%%%%%%%

\maketitle

%\begin{history}
%\received{Day Month Year}
%\revised{Day Month Year}
%\end{history}

%\ccode{PACS numbers:  }
\section{Introduction}

The charmless hadronic B decays are important in determining the CKM
phase angle and searching for new physics signal in B physics. The
theory of hadronic B decays involving non-perturbative dynamics is
improving fast since the so called naive factorization approach
\cite{bsw,9804363}. In recent years, the QCD factorization approach
(QCDF) \cite{qcdf} and perturbative QCD factorization (pQCD)
approach \cite{pqcd} together with the soft-collinear effective
theory \cite{scet} solved a lot of problems in the non-leptonic  B
meson decays. Although most of the branching ratios measured by the
B factory experiments can be explained by any of the theories, the
direct CP asymmetries measured by the experiments are ever predicted
with the right sign only by the pQCD approach \cite{direct}. The
LHCb experiment will soon run in the end of 2008. With a very large
luminosity,  it will accumulate a lot of B and  $B_s$ events. The
progress in both theory and experiment encourages us to apply the
pQCD approach to the charmless $B_s$ decays \cite{PQCDBs}.

In the hadronic $B(B_s)$ decays, there are various energy scales
involved. The factorization theorem allows us to calculate them
separately. First, the physics from the electroweak scale down to b
quark mass scale is described by the renormalization group running
of the Wilson coefficients of effective four quark operators.
Secondly, the hard scale from b quark mass scale to the
factorization scale $\sqrt{\Lambda m_B}$ are calculated by the hard
part calculation in the perturbative QCD approach \cite{li2003}.
When doing the integration of the momentum fraction $x$ of the light
quark, end point singularity will appear in the collinear
factorization (QCDF and SCET) which breaks down the factorization
theorem. In the pQCD approach, we do not neglect the transverse
momentum $k_T$ of the light quarks in meson. Therefore the endpoint
singularity disappears. The inclusion of transverse momentum will
also give large double logarithms ln$^2k_T$ and ln$^2x$ in the hard
part calculations. Using the renormalization group equation, we can
resum them for all loops to the leading order resulting Sudakov
factors. The Sudakov factors suppress the endpoint contributions to
make the pQCD calculation consistent\cite{pqcd}.

\begin{figure}
 \includegraphics[width=0.4\textwidth]{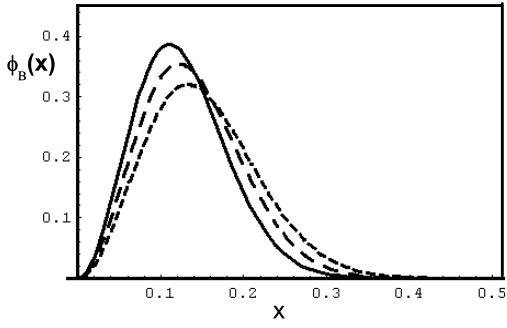} \caption{ $B_s$
meson distribution amplitudes. The solid-, dashed-, and tiny-dashed-
lines correspond to $\omega_B=0.45~\mathrm{GeV}$,
$0.5~\mathrm{GeV}$,  and $0.55~\mathrm{GeV}$.}\label{bswave}
 \end{figure}

 The physics below the
factorization scale is non-perturbative in nature, which is
described by the hadronic wave functions of mesons. They are not
perturbatively calculable, but universal for all the decay
processes. Since many of the hadronic
 and semi-leptonic $B$ decays have been measured well in the two B
 factory experiments, the light wave functions are strictly
 constrained.
  Therefore, it is useful to use the same light meson wave functions in our  $B_s$
decays determined from the hadronic $B$ decays. The uncertainty of
the hadronic wave functions will come mainly from the SU(3) breaking
effect between the $B_s$ wave function and $B$ wave function
\cite{PQCDBs}.
 In recent years, a lot of studies have been
performed for the $B_d^0$ and $B^\pm$ decays  in the pQCD approach
\cite{pqcd}. The parameter $\omega_b=0.40~\mathrm{GeV}$  has been
fixed there using the rich experimental data on the $B_d^0$ and
$B^\pm$ mesons. In the SU(3) limit, this parameter should be the
same in $B_s$ decays. Considering  a small SU(3) breaking, the $s$
quark momentum fraction here should be  a little larger than that of
the $u$ or $d$ quark in the $B$ mesons, since  the
  $s$ quark is heavier than the $u$ or $d$ quark. The shape
of the distribution amplitude is shown in Fig.\ref{bswave} for
$\omega_B=0.45~\mathrm{GeV}$, $0.5~\mathrm{GeV}$, and
$0.55~\mathrm{GeV}$. It is easy to see that the larger $\omega_b$
gives a larger momentum fraction to the $s$ quark.  We will use
$\omega_b=0.50\pm 0.05~\mathrm{GeV}$ in this paper for the $B_s$
decays, which characterize the fact that the  $s$ quark in $B_s$
meson carries a littler larger momentum fraction than the $d$ quark
in the $B_d$ meson.

 \section{Results and Discussion}

For $B$($B_s$) meson decays with two light  mesons in the final
states, the light mesons obtain large momentum of 2.6GeV in the
$B$($B_s$) meson rest frame. All the quarks inside the light mesons
are therefore energetic and collinear like. Since the heavy b quark
in $B$($B_s$) meson carries most of the energy of $B$($B_s$) meson,
the light quark in $B$($B_s$) meson is soft. In the usual emission
diagram of $B$($B_s$) decays, this quark goes to the final state
meson without electroweak interaction with other quarks, which is
called a spectator quark. Therefore there must be a connecting hard
gluon to make it from soft like to collinear like.
   The hard part of the interaction
becomes six quark operator rather than four. The soft dynamics here
is included in the meson wave functions. The decay amplitude is
infrared safe and can be factorized as the following formalism:
\begin{equation}
C(t) \times H(t) \times \Phi (x) \times \exp\left[ -s(P,b) -2 \int
_{1/b}^t \frac{ d \bar\mu}{\bar \mu} \gamma_q (\alpha_s (\bar \mu))
\right], \label{eq:factorization_formula}
\end{equation}
where $C(t)$ are the corresponding Wilson coefficients of four quark
operators; $\Phi (x)$ are the meson wave functions and the
factorization scale $t$ denotes the largest energy scale of hard
process $H$, which is the typical energy scale in pQCD approach. The
exponential of $S$ function is the so-called Sudakov form factor
resulting from the resummation of double logarithms occurred in the
QCD loop corrections, which can suppress the contribution from the
non-perturbative region.

 First we give the numerical results in the
pQCD approach for the form factors at maximal recoil. For the
$B_{d(s)}\to P$ form factors, we obtain:
\begin{eqnarray}
F_{0}^{B\to\pi}= 0.29^{+0.07+0.00}_{-0.05-0.01},&& F_{0}^{B_s\to
K}=0.28^{+0.05+0.00}_{-0.04-0.01},\nonumber\\
F_{0}^{B\to \eta_n}=0.28^{+0.06+0.00}_{-0.05-0.00},&& F_{0}^{B\to
K}=0.37^{+0.09+0.07}_{-0.01-0.00},\nonumber\\
 F_{0}^{B_s\to\eta_s}=0.42^{+0.08+0.01}_{-0.04-0.01}.\label{form1}
\end{eqnarray}
For the $B_{d(s)}\to V$ form factors, we obtain:
\begin{equation}
\begin{array}{ll}
 V^{B \rightarrow K^{\ast}}=0.31^{+0.06+0.01}_{-0.05-0.00}, &  A_0^{B \rightarrow
K^{\ast}}=0.37^{+0.07+0.01}_{-0.06-0.01},\\
 V^{B
\rightarrow \rho}=0.26^{+0.05+0.01}_{-0.04-0.00}, &
 A_0^{B \rightarrow \rho}=0.32^{+0.06+0.00}_{-0.05-0.01},
 \\
 V^{B \rightarrow \omega}=0.24^{+0.05+0.01}_{-0.04-0.00}, &  A_0^{B \rightarrow
\omega}=0.29^{+0.05+0.00}_{-0.05-0.01}\\
V^{B_s \rightarrow K^{\ast}}=0.27^{+0.04+0.00}_{-0.04-0.01},&
A_0^{B_s \rightarrow
K^{\ast}}=0.33^{+0.05+0.00}_{-0.05-0.01},  \\
 V^{B_s
\rightarrow \phi}=0.32^{+0.05+0.00}_{-0.04-0.01},&  A_0^{B_s
\rightarrow \phi}=0.38^{+0.07+0.01}_{-0.05-0.00}, \\
 A_1^{B_s \rightarrow \phi}=0.24^{+0.04+0.00}_{-0.03-0.01},
 & A_1^{B \rightarrow K^{\ast}}=0.24^{+0.05+0.00}_{-0.04-0.00}, \\
 A_1^{B \rightarrow \rho}=0.21^{+0.04+0.00}_{-0.03-0.00}, & A_1^{B
\rightarrow \omega}=0.19^{+0.04+0.01}_{-0.03-0.00},\\
 A_1^{B_s
\rightarrow K^{\ast}}=0.21^{+0.03+0.00}_{-0.03-0.01},
\end{array}\label{form2}
\end{equation}
where $f_B=0.21 \pm 0.02\mbox{ GeV}$, $\omega_B=0.40\mbox{ GeV}$
(for the $B^\pm$ and $B_d^0$ mesons) and $f_{B_s}=0.26 \pm 0.02$
GeV, $\omega_{B_s}=0.50 \pm 0.05$ GeV (for the $B_s^0$ meson). They
quantify the SU(3)-symmetry breaking effects in the form factors in
the pQCD approach.
 From these form factors in eq.(\ref{form1},\ref{form2}), one can notice
 that the $B \to K^{(*)}$ transition form factors are larger than
 that of $B \to \pi(\rho)$ due to the SU(3) breaking effect. Usually
 the same kind of SU(3) breaking effect is also expected between the
 $B \to \pi(\rho)$   and $B_s \to K^{(*)}$  transition form factors,
 but the numbers shown in eq.(\ref{form1},\ref{form2}) are quite similar.
 This is due to the fact that the effect of a larger $f_{B_s}$ decay constant
canceled by a larger $\omega_{B_s}$ parameter of the $B_s$ wave
function.

 The numerical results of the $B(B_s)$ decays
branching ratios and CP asymmetry parameters  are displayed in
Ref.~\cite{pqcd,PQCDBs}. Most of the $B$ decay channels are measured
by B factories; while  for charmless $B_s$ decays, only several are
measured by the CDF collaboration\cite{CDFBsKKrecent}. The measured
branching ratios of $B$ and $B_s$ decays are
  consistent with the theoretical calculations.
   The calculated branching ratios from the three kinds of
methods overlap with each other, considering the still large
theoretical and experimental uncertainties.

In table~\ref{tab:exp}, the only measured CP asymmetry in $B_s \to
K^- \pi^+$ decay prefer our pQCD approach rather than QCDF approach.
This is similar with the situation in $B$ decays. The direct CP
asymmetry is proportional to the sine of the strong phase difference
of two  decay topologies \cite{direct}. The strong phase in our pQCD
approach is mainly from the chirally enhanced space-like penguin
diagram (annihilation penguin); while in the QCDF approach, the
strong phase mainly comes from the virtual charm quark loop
diagrams. The different origin of strong phases gives different sign
to the direct CP asymmetry imply a fact that the dominant strong
phase in the charmless decays should come from the annihilation
diagrams.  It should be noted that the SCET approach can not predict
the direct CP asymmetry of $B$ decays directly, since they need more
experimental measurements as input. However, it also gives the right
CP asymmetry for $B_s$ decay if with the input of experimental CP
asymmetries of $B$ decays, which means good SU(3) symmetry here
\cite{SCETBs}. Unlike pQCD approach and QCDF approach, in SCET, the
main strong phase responsible for the direct CP asymmetry is from
the non-perturbative charming penguins. Although it has the same
topology as the annihilation penguin, it may give different mixing
induced CP asymmetry for neutral $B(B_s)$ decays \cite{SCETBs}.

\begin{table}[t]
\caption{The branching ratios (  $10^{-6}$) and CP asymmetry (\%)
calculated in pQCD approach, QCDF
 and SCET approaches together with Experimental Data.\label{tab:exp}} \vspace{0.1cm}
\begin{center}
\begin{tabular}{|c|c|c|c|l|}
\hline & SCET  &     QCDF   &   pQCD    &   EXP
\\ \hline
$B(B_s\to K^- \pi^+) $ &  $4.9\pm 1.8$  & $  10\pm6 $& $ 11 \pm 6$
& $5.0\pm 1.3$\\
$B( B_s\to K^- K^+) $ &$18\pm 7$  &    $ 23\pm 27$ & $17 \pm9$&
$24 \pm 5 $ \\
$B(B_s\to \phi \phi) $ & & $22 \pm30$&$ 33 \pm13$&$ 14 \pm 8$
 \\
\hline $A_{CP}(B_s\to K^- \pi^+)$  &$20 \pm 26$ & $     -6.7 \pm16$&
$ 30 \pm 6$&$    39 \pm17$
\\ \hline
\end{tabular}
\end{center}
\end{table}

For the $B_s\to PP$ branching ratios, the results for QCDF and pQCD
approaches are quite similar, which also happens in the $B\to PP$
decays. The large chiral enhancement factor is adopted in QCDF to
give a large penguin contribution here, while in pQCD approach, an
additional chirally enhanced penguin annihilation amplitude is
included, which gives the same order effect. However, for the $B_s
\to PV$ channels, the QCDF results are systematically smaller than
the pQCD results, which is again happens in the $B\to PV$ channels,
where pQCD results are more favored by B factory experiments. Since
there is no more chiral enhancement factor here in the emission type
penguin diagram, the QCDF results can not catch up the experiments
and the pQCD results with penguin annihilation contributions.

In the $B \to VV$ decays, again, it is the annihilation penguin
diagram gives large transverse polarization contribution to the
penguin dominant B decays \cite{vv}.  For the $B_s \to VV$ decays,
we also give the polarization fractions in addition to the branching
ratios and CP asymmetry parameters \cite{PQCDBs}. Similar to the
$B\to VV$ decay channels, we also have large transverse polarization
fractions for the penguin dominant processes, such as $B_s \to
\phi\phi$, $B_s\to K^{*+}K^{*-}$, $K^{*0}\bar K^{*0}$ decays, whose
transverse polarization fraction can reach 40-50\%.

\section{SU(3) breaking effect}

\begin{table*}
\caption{\label{Amplitudes}Contributions from the various topologies
to the decay amplitudes (squared) for the four indicated decays.
Here, $\mathcal{T}$ is the contribution from the color favored
emission diagrams; $\mathcal{P}$ is the penguin contribution from
the emission diagrams; $\mathcal{E}$ is the contribution from the
W-exchange diagrams; $\mathcal{P_A}$ is the contribution from the
penguin annihilation amplitudes; and $\mathcal{P_{EW}}$ is the
contribution from the electro-weak penguin induced amplitude.}
\begin{ruledtabular}
\begin{tabular}{c|ccccc}
  %\hline
 mode ($\mbox{GeV}^2$)& $|\mathcal{T}|^2$&$|\mathcal{P}|^2$&$ |\mathcal{E}|^2$&$|\mathcal{P_A}|^2$&$|\mathcal{P_{EW}}|^2$\\ \hline
   $B_d \to \pi^+\pi^-$   &~~~$1.5$~~~&  $9.2\times 10^{-3}$  &$6.4\times 10^{-3}$ &$7.5\times 10^{-3}$  & $2.7\times 10^{-6}$ \\

   $B_s \to \pi^+ K^- $   &~~~$1.4$~~~&  $7.4\times 10^{-3}$  & $0$                &$7.0\times 10^{-3}$  & $5.4\times 10^{-6}$
   \\ \hline\hline
   $B_d \to K^+\pi^-  $   &~~~$2.2$~~~&  $18.8\times 10^{-3}$   &  0                 &$4.7\times 10^{-3}$  & $7.4\times 10^{-6}$  \\
   $B_s \to K^+K^-    $   &~~~$2.0$~~~&  $14.7\times 10^{-3}$  &
$4.6\times 10^{-3}$& $9.8\times 10^{-3}$ & $3.1\times 10^{-6}$  \\
   \end{tabular}
\end{ruledtabular}
\end{table*}

 The SU(3) breaking effect comes
mainly from the $B_s(B_d)$ meson decay constant and distribution
amplitude parameter, light meson decay constant and wave function
difference, and various decay topology differences. As an example we
mainly focus on  the decays $B \to \pi \pi$, $B \to K \pi$, $B_s \to
K\pi$ and $B_s \to KK$, as they can be related by SU(3)-symmetry. A
question of considerable interest is the amount of SU(3)-breaking in
various topologies (diagrams) contributing to these decays. For this
purpose, we present in Table~\ref{Amplitudes} the magnitude of the
decay amplitudes (squared, in units of GeV$^2$) involving the
distinct topologies for the four decays modes. The first two decays
in  this table are related by U-spin symmetry $(d \to s)$ (likewise
the two decays in the lower half). We note that the assumption of
U-spin symmetry for the (dominant) tree ($\mathcal{T}$) and penguin
($\mathcal{P}$)
 amplitudes in
the emission diagrams is quite good, it is less so in the other
topologies, including the contributions from the $W$-exchange
diagrams, denoted by $\mathcal{E}$ for which there are non-zero
contributions for the flavor-diagonal states $\pi^+\pi^-$ and
$K^+K^-$ only. The U-spin breaking is large
 in the electroweak penguin induced amplitudes $\mathcal{P_{EW}}$,
and in the penguin annihilation amplitudes $\mathcal{P_A}$ relating
the decays $B_d \to K^+\pi^-$ and $B_s \to K^+ K^-$.
 In the SM, however,
the amplitudes $\mathcal{P_{EW}}$ are negligibly small.

 In $\overline{B_d^0}\to K^-\pi^+$ and $\overline{B_s^0}\to
K^+\pi^-$, the branching ratios are very different from each other
due to the differing strong and weak phases entering in the tree and
penguin amplitudes. However, as shown by He and
Gronau~\cite{Gronau:2000zy}, the two relevant products of the CKM
matrix elements entering in the expressions for the direct CP
asymmetries in these decays are equal, and, as stressed by
Lipkin~\cite{KpiLipkin} subsequently, the final states in these
decays are charge conjugates, and the strong interactions being
charge-conjugation invariant, the direct CP asymmetry in
$\overline{B_s^0}\to K^-\pi^+$ can be related to the well-measured
CP asymmetry in the decay $\overline{B_d^0}\to K^+\pi^-$ using
U-spin symmetry.

\begin{figure}
\includegraphics[width=0.4\textwidth]{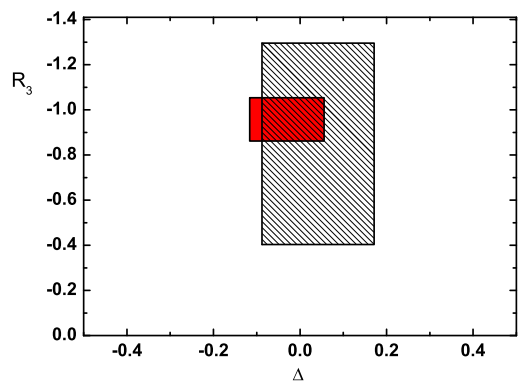}
\caption{\label{R3} $R_3$ vs $\Delta$: The red (smaller) rectangle
is the pQCD estimates worked out in this paper. The experimental
results with their $\pm 1 \sigma$ errors are shown as the larger
 rectangle.}
\end{figure}

Following the suggestions in the literature, we can  define the
following two parameters:
\begin{eqnarray}
R_3&\equiv&\frac{|A(B_s\to\pi^+K^-)|^2-|A(\bar
B_s\to\pi^-K^+)|^2}{|A(B_d\to\pi^-K^+)|^2-|A(\bar
B_d\to\pi^+K^-)|^2},\\
\Delta&=&\frac{A^{dir}_{CP}(\bar B_d\to\pi^+K^-)}{A^{dir}_{CP}(\bar
B_s\to\pi^-K^+)}+\frac{BR(B_s\to\pi^+K^-)}{BR(\bar
B_d\to\pi^+K^-)}\cdot\frac{\tau(B_d)}{\tau(B_s)}.\nonumber\\
\end{eqnarray}
The standard model predicts $R_3=-1$ and $\Delta=0$ if we assume
$U$-spin symmetry. Since we have a detailed dynamical theory to
study the SU(3) (and U-spin) symmetry violation, we can check in
pQCD approach how good quantitatively this symmetry is in the ratios
$R_3$ and $\Delta$. We
  get $R_3=-0.96 ^{+0.11}_{-0.09}$ and $\Delta =
-0.03\pm 0.08$. Thus, we find that these quantities are quite
reliably calculable, as anticipated on theoretical grounds. SU(3)
breaking and theoretical uncertainties are very small here, because
most of the breaking effects and uncertainties are canceled due to
the definition of $R_3$ and $\Delta$. On the experimental side, the
results for $R_3$ and $\Delta$ are:~\cite{CDFBsKKrecent}
\begin{equation}
R_3=-0.84\pm0.42\pm0.15,\;\; \Delta=0.04\pm0.11\pm0.08.
\end{equation}
We conclude that SM is in  good agreement with the data, as can also
be seen in Fig.~\ref{R3} where we plot theoretical predictions for
$R_3$ vs.~$\Delta$ and compare them with the current measurements of
the same. The  measurements of these quantities are rather imprecise
at present,
 a situation which we hope will greatly improve at the LHCb experiment.

\section{Summary}

 Based on
the $k_T$ factorization, pQCD approach is infrared safe. Its
predictions on the branching ratios and CP asymmetries of the
$B^0(B^\pm)$ decays are tested well by the B factory experiments.
Using those tested parameters from these decays,  we predict
branching ratios and CP asymmetries of a number of charmless decay
channels $B_s\to PP$, $PV$ and $VV$ in the perturbative QCD
approach. The experimental measurements of the three $B_s$ decay
channels are consistent with our numerical results. Especially the
measured direct CP asymmetry of $B_s\to \pi^-K^+$ agree with our
calculations. We also discuss the SU(3) breaking effect in these
decays, which is at least around 20-30\%. We also show that the
He-Gronau-Lipkin sum rule works quite well in the standard model,
where the SU(3) breaking effects mainly cancel.

The annihilation penguin diagram is chirally enhanced in the pQCD
approach. Its large contribution and strong phase give the right
sign for direct CP asymmetry, the large branching ratios for the
$B\to PV $ decays and large transverse polarization fraction for the
$B\to VV $ decays. Its important role is also realized later in the
QCDF approach.

\begin{acknowledgments}
We are grateful to the collaborators of this work: A. Ali, G.
Kramer, Y. Li, Y.L. Shen, W. Wang and Y.M. Wang. This Work is
supported by National Science Foundation of China under Grant
No.10735080 and 10625525.
\end{acknowledgments}

\bigskip % extra skip inserted

\end{document}